\documentclass[10pt]{article}
\usepackage{OSAmeet,graphicx,amsmath,amssymb,cite,subfig}
\begin{document}

\topmargin-0.75in

\title{An Anisotropic Metamaterial Leaky Waveguide}

\author{Huikan Liu and Kevin J. Webb}
\vspace{2mm}
\address{School of Electrical and Computer Engineering, Purdue
  University, West Lafayette, IN 47907, USA}
\vspace{2mm}
\email{Email: webb@purdue.edu}


\graphicspath{{figures/}}

\begin{abstract}
We propose a leaky optical waveguide achieved with a uniaxially anisotropic
metamaterial that supports both forward and backward leaky
waves. The backward leaky nature is exploited in a sub-diffraction imaging system.
\end{abstract}
\vspace{2mm} 

Leaky wave antennas can offer high directivity and frequency
scanning \cite{Oliner_LeakyWaves}. An isotropic dielectric slab with
refractive index $n>1$, which supports guided modes, does not
radiate effectively unless spatial variations are introduced to
excite fast-wave space harmonics \cite{Oliner_LeakyWaves}.
Translating such components to nanophotonics presents fabrication
difficulties. To realize uniform leaky wave structures, potentially
in the optical domain, a metamaterial slab with $n\ll1$ that leads
to enhanced radiation directivity has been proposed and investigated
\cite{Enoch_PRL_2002, Baccarelli_IEEETMTT_2005}. This approach
suffers the limitation of large transverse slab dimensions, because
the operating wavelength inside the slab is much greater than the
free space wavelength due to the small value of $n$
\cite{Alu_IEEETAP_2007}. Low refractive index metamaterials that
require small material permeabilities face challenges of
implementation \cite{Baccarelli_IEEETMTT_2005}. In addition,
radiation occurs in a narrow angular range between broadside and the
critical angle $\sin^{-1}(n)$, which presents a restriction for beam
scanning. Finally, we note that while a microstrip transmission line
loaded with series capacitors and shunt inductors has been studied
to realize metamaterial leaky-wave antennas in the microwave
frequency regime \cite{Itoh_Metamaterials}, identifying the optical
counterparts of shunt inductors and series capacitors is difficult.
To overcome these limitations, we present a forward/backward
radiating leaky waveguide at optical wavelengths based on
non-magnetic uniaxially anisotropic metamaterials having
permittivity tensors in which one component is different from the
others in sign \cite{Smith_PRL_2003}. The resulting hyperbolic
dispersion relation enables radiation from backfire ($-90^{\circ}$)
to endfire ($90^{\circ}$), and the slab thickness can be small
relative to the free space wavelength.

\begin{figure}[!htb]
\centering
\subfloat[\footnotesize]{\label{fig:Schematic}\includegraphics[width=0.3\textwidth]{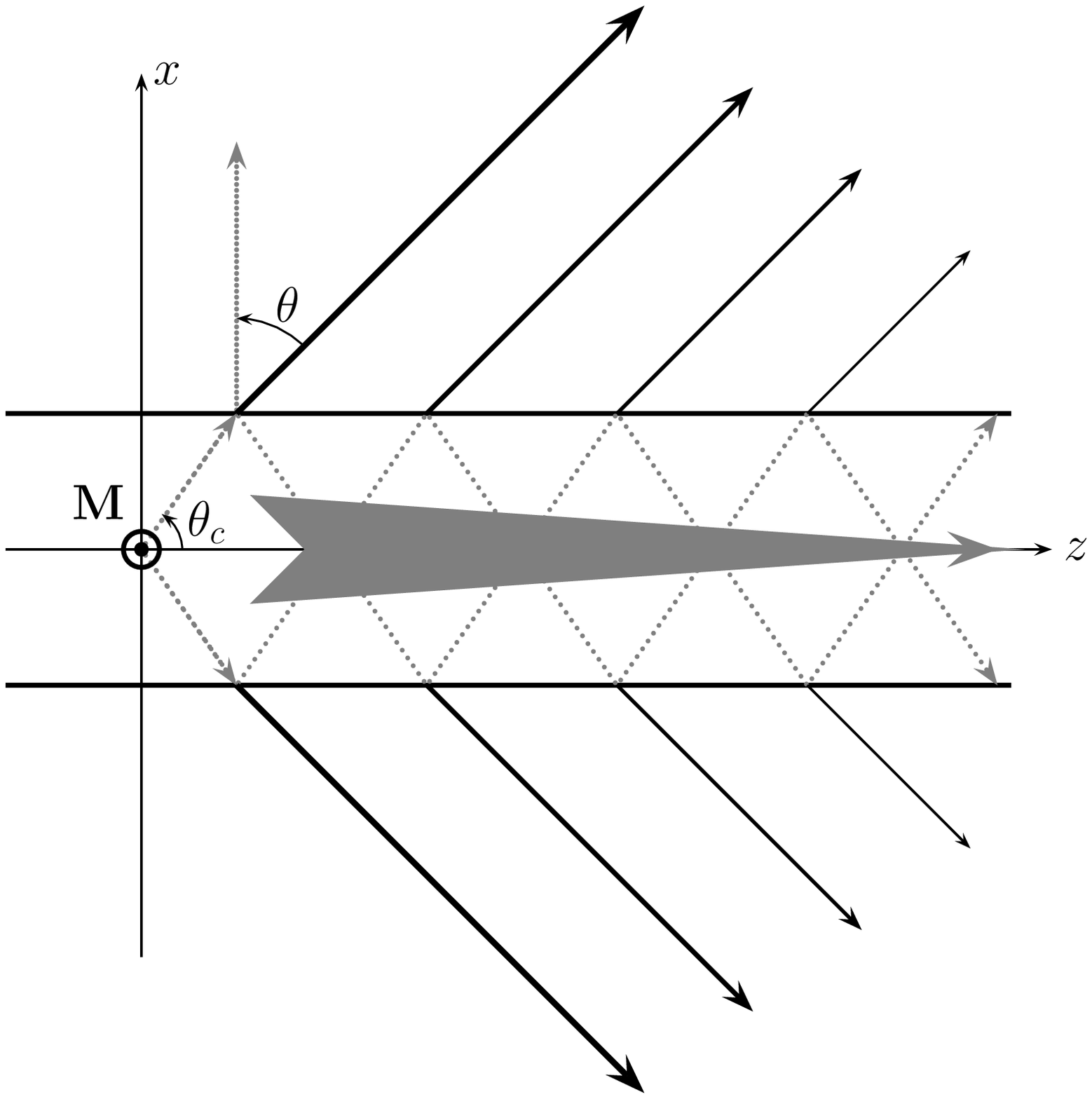}}\hfil
\subfloat[\footnotesize]{\label{fig:FHField}\includegraphics[width=0.35\textwidth]{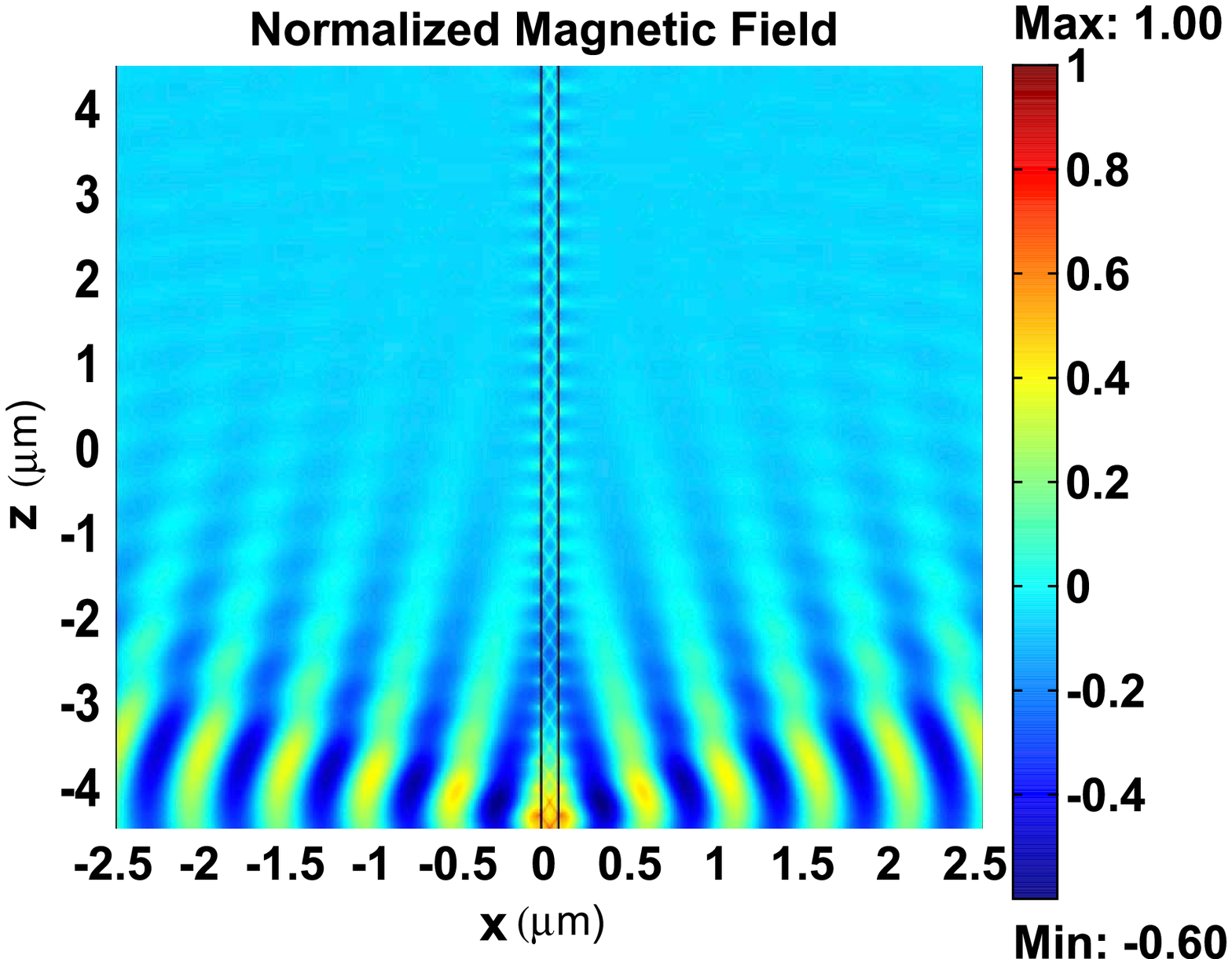}}\hspace{-1em}
\subfloat[\footnotesize]{\label{fig:BHField}\includegraphics[width=0.35\textwidth]{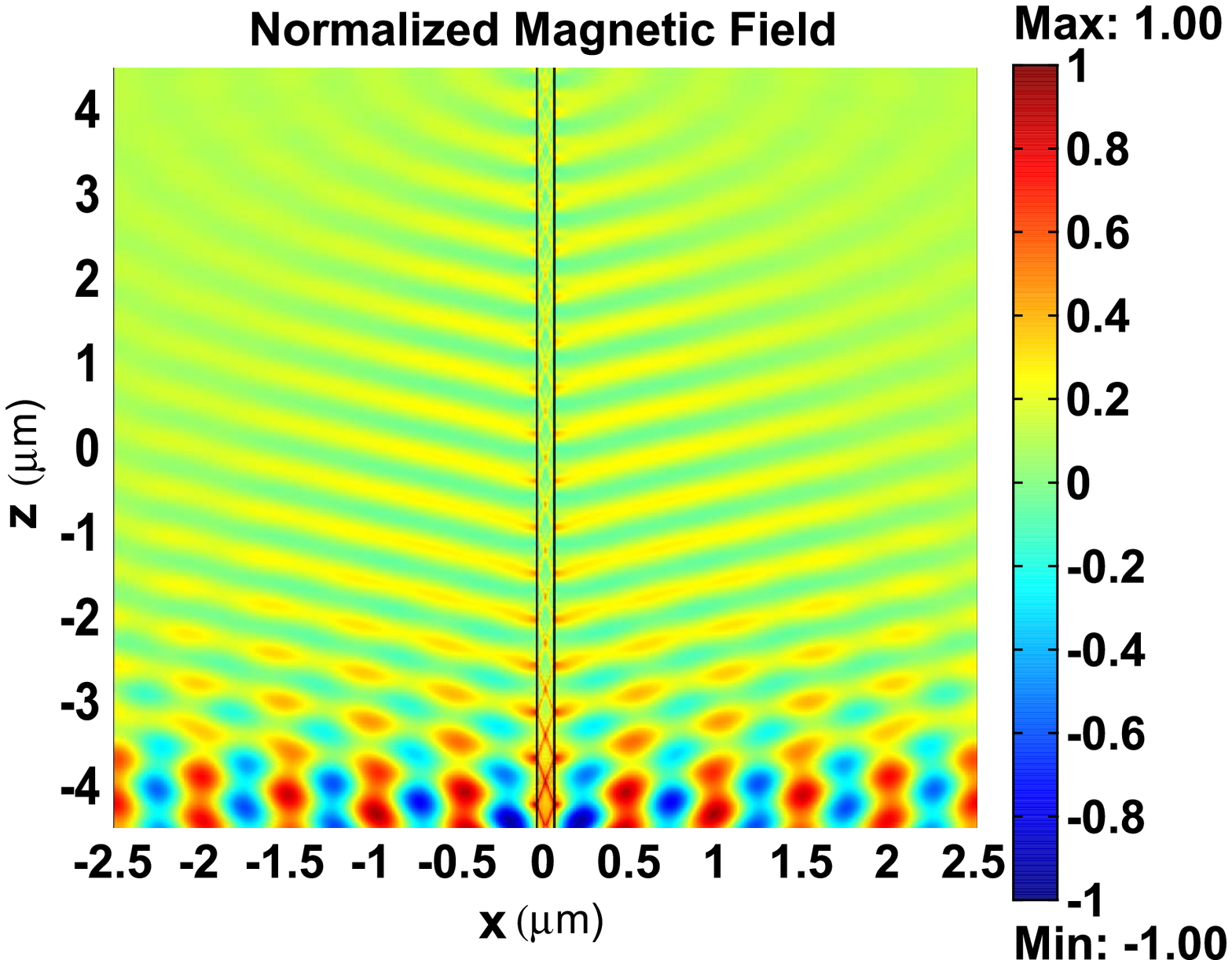}}\\
\vspace{-1em}\caption{\subref{fig:Schematic} Schematic
representation of an anisotropic leaky wave structure. The width of the slab is
$0.1\,\mu\textrm{m}$, and $\lambda=0.5\,\mu\textrm{m}$.
\subref{fig:FHField} Full-wave (finite element method) simulated
$H$-field distribution of a forward anisotropic leaky waveguide with
$\epsilon_x=0.01$, $\epsilon_z=-0.1$. A
filamentary magnetic current line source is located at
$(x=0,\,z=-4.5\,\mu\textrm{m})$. \subref{fig:BHField} Backward
anisotropic leaky waveguide with $\epsilon_x=-0.01$,
$\epsilon_z=0.3$.}\label{fig:LeakyWaveguide}
\end{figure}
We consider the anisotropic waveguide configuration of
Fig.~\ref{fig:LeakyWaveguide}\subref{fig:Schematic}, where a
transverse magnetic (TM) wave ($H_y,E_x,E_z$) is excited by a
magnetic line current source $\mathbf{M}$ located at $x=0$ and
travels in the $+z$ direction with a propagation constant
$k_z=\beta_z + i\alpha_z$. This excitation generates all the even
modes of the structure, in terms of $H_y$. The anisotropic slab can
be characterized by a dielectric tensor $\boldsymbol{\epsilon}={\hat
x}{\hat x}\epsilon_x + ({\hat y}{\hat y} + {\hat z}{\hat
z})\epsilon_z$, with $\epsilon_{x}\epsilon_{z}<0$, and the
surrounding medium is assumed to be free space. The TM fields in the
anisotropic slab have a dispersion relation $k_x^2 \epsilon_{z}
^{-1} +k_{z}^2 \epsilon_{x} ^{-1} = k_0^2$, and in free space,
$k_{0x}^2 +k_{z}^2 = k_0^2$. Leaky modes of the structure are not
confined to the slab, i.e., $k_{0x}$ is approximately real, which
leads to the requirement that $-k_0<\beta_z<k_0$ and
$\alpha_z\ll|\beta_z|$ \cite{Oliner_LeakyWaves}. The radiation
occurs at angle $\theta = \sin^{-1}(\beta_z/k_0)$ with respect to
the broadside direction. In the regime $k_x^2\gg |\epsilon_z|
k_0^2$, the longitudinal wavenumber inside the anisotropic slab can
be approximately expressed as $k_{z} \approx
\pm\sqrt{-{\epsilon_{x}}/{\epsilon_{z}}}k_x$, where the sign depends
on the signs of $\epsilon_{x}$ and $\epsilon_{z}$, namely, ``$+$" is
assumed if $\epsilon_{x}>0$ and $\epsilon_{z}<0$, whereas
$\epsilon_{x}<0$ and $\epsilon_{z}>0$ leads to ``$-$"
\cite{Liu_OL_2009}. Numerical analysis reveals that the slab is
capable of supporting a fast wave ($|\beta_z|<k_0$) and leaking
power along the waveguide given appropriate values of $k_x$. A
mathematically sufficient condition for the dielectric tensor for
leakage into free space is that $|\epsilon_{x}|\ll 1$, which is a
salient sign of anisotropic $\epsilon$-near-zero metamaterials
\cite{Engheta_Science_2007}. If $\epsilon_{x}>0$ and
$\epsilon_{z}<0$, the structure leaks forward; if $\epsilon_{x}<0$
and $\epsilon_{z}>0$, the structure radiates backward. The
forward/backward propagation characteristics are direct consequences
of the right/left-handedness of strongly anisotropic media
\cite{Podolskiy_PRB_2005}. In
Figs.~\ref{fig:LeakyWaveguide}\subref{fig:FHField} and
\ref{fig:LeakyWaveguide}\subref{fig:BHField}, we show the field
distributions of forward and backward leaky waveguides with
sub\-wavelength thickness ($=0.1\,\mu\textrm{m}=\lambda/5$). In
Fig.~\ref{fig:LeakyWaveguide}\subref{fig:FHField}, the supported
mode has $k_z/k_0 = 0.12 + i0.065$, by solving a transverse resonance
relation \cite{Oliner_LeakyWaves}, and leads to the radiation angle
$\theta = 7.0^{\circ}$. In
Fig.~\ref{fig:LeakyWaveguide}\subref{fig:BHField}, the dominant mode
has $k_z/k_0 = -0.91 + i0.0074$, resulting in a beam angle of $\theta =
-65.2^{\circ}$.

\begin{figure}[!htb]
\vspace{-1em}\centering
\subfloat[\footnotesize]{\label{fig:SiO2.Ag_563.6nm}\includegraphics[width=0.34\textwidth]{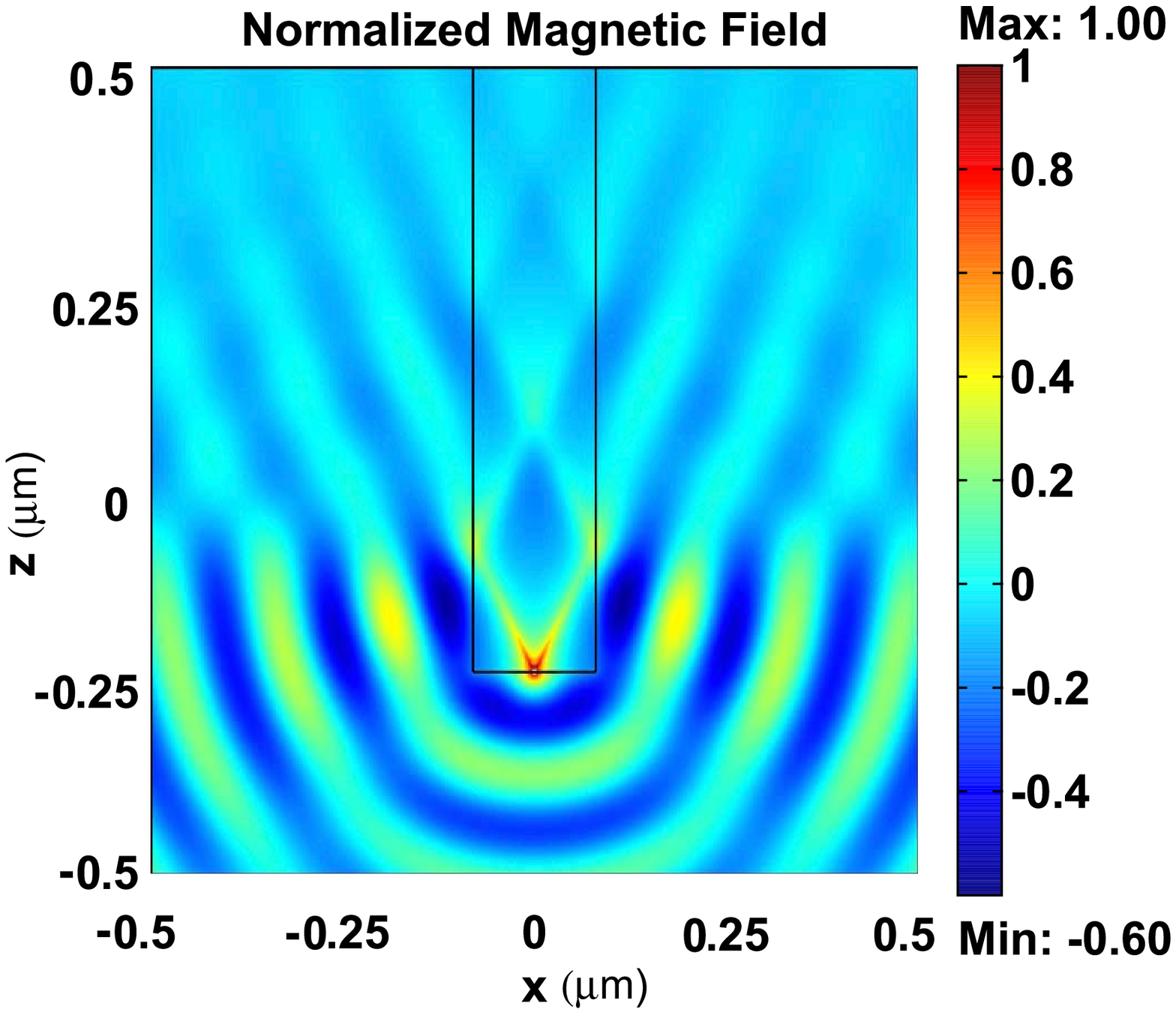}}\hfil
\subfloat[\footnotesize]{\label{fig:SiO2.Ag_563.6nm_Focusing}\includegraphics[width=0.35\textwidth]{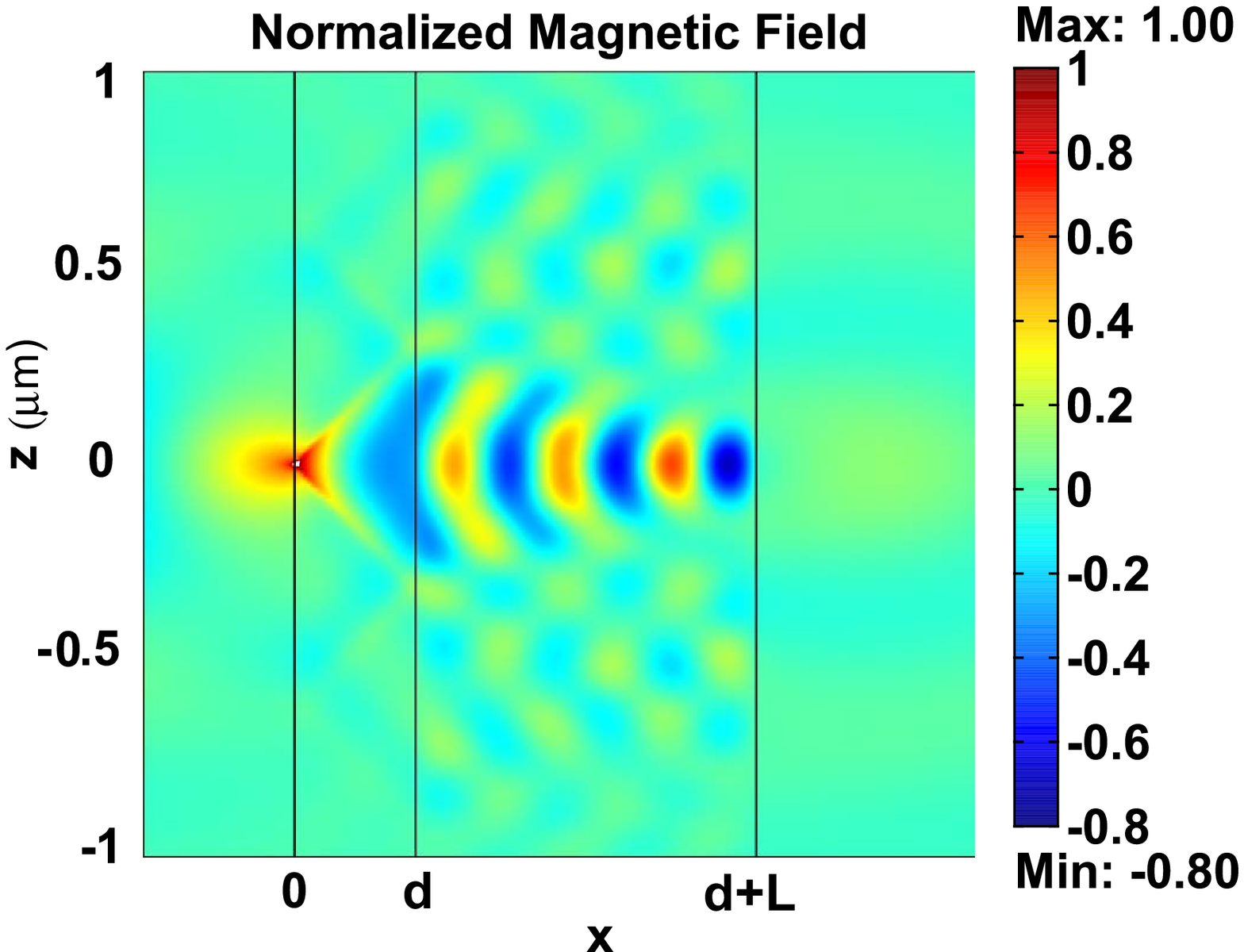}}\hfil
\subfloat[\footnotesize]{\label{fig:SiO2.Ag_LensImage}\includegraphics[width=0.3\textwidth]{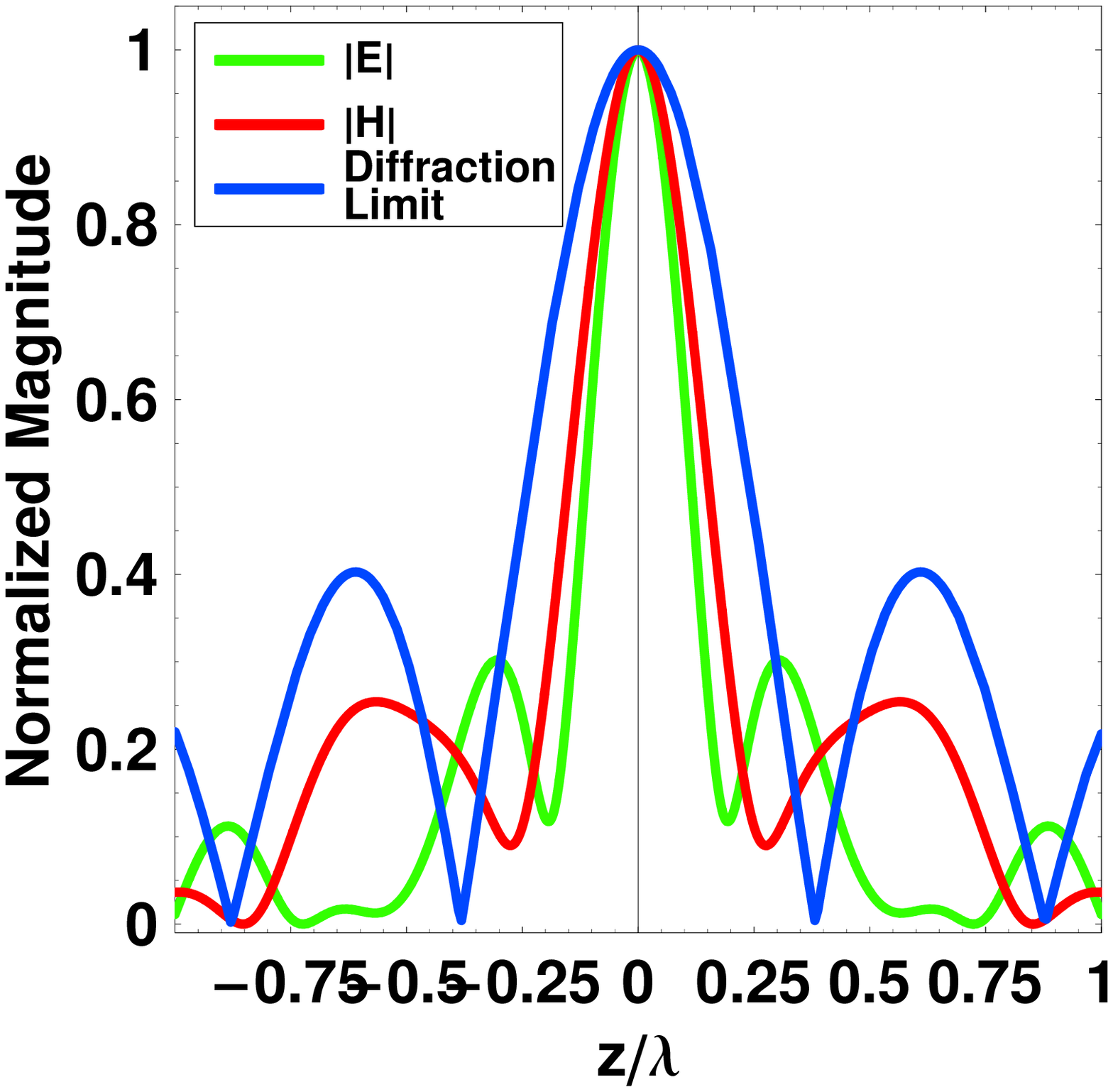}}\\
\vspace{-1em} \caption{\subref{fig:SiO2.Ag_563.6nm} $H$-field
distribution of a Ag/$\textrm{SiO}_2$ bulk anisotropic leaky
waveguide embedded in a Si host. The volume fraction for Ag is
$0.2$. The width of the slab is $0.16\,\mu\textrm{m}$. A magnetic
current line source is located at $(x=0,\,z=-0.25\,\mu\textrm{m})$.
Operating wavelength $\lambda=0.5636\,\mu\textrm{m}$, $\epsilon_{\text{Si}}= 16.3+
i0.26$, $\epsilon_{\text{Ag}}=-11.89+i0.83$,
$\epsilon_{\textrm{SiO}_2}=2.13$ \cite{Palik_Optical}, giving
$\epsilon_x=-0.67+i0.17$, $\epsilon_z=2.79+i9.05\times10^{-3}$.
\subref{fig:SiO2.Ag_563.6nm_Focusing} A Ag/$\textrm{SiO}_2$ bulk
anisotropic leaky waveguide juxtaposed with a Si slab. The thickness
of the waveguide is $d=0.16\,\mu\textrm{m}$, the thickness of silicon
slab is $L=0.45\,\mu\textrm{m}$, and the surrounding medium is free
space. The excitation current line source is located at $x=0$.
\subref{fig:SiO2.Ag_LensImage} The field profile at the image plane
$x=d+L$ with respect to the operating wavelength.
}\label{fig:RealWaveguide}
\end{figure}
One way to demonstrate an optical leaky waveguide with anisotropic
metamaterials with enhanced local radiation effects based on use of
a metal-insulator stack is to embed the slab in a dielectric host
with large permittivity $\epsilon_r$. In
Fig.~\ref{fig:RealWaveguide}\subref{fig:SiO2.Ag_563.6nm}, we
investigate the homogenized Ag/$\textrm{SiO}_2$ metal-insulator
stack embedded in a silicon (Si) host as a possible realization of
an anisotropic leaky waveguide. The finite element simulation result
clearly demonstrates that the slab behaves as an optical leaky
waveguide radiating backwards, due to the signs of the effective
$\epsilon_x$ and $\epsilon_z$, and the dominant $k_z$ is obtained as
$k_z/k_0 = -3.40 + i0.45$. Since the leaky mode supported by the
system can have
$n_{\textrm{Si}}k_0>|\mathop{\mathrm{Re}}(k_z)|>k_0$, a planar
near-field imaging system that transfers sub-diffraction features
can be envisioned by pairing a backward leaky waveguide with a
silicon slab, as shown in
Figs.~\ref{fig:RealWaveguide}\subref{fig:SiO2.Ag_563.6nm_Focusing}
and \ref{fig:RealWaveguide}\subref{fig:SiO2.Ag_LensImage}. As
illustrated in
Fig.~\ref{fig:RealWaveguide}\subref{fig:SiO2.Ag_563.6nm_Focusing},
the waves radiated from the source emerge as strongly confined beams
inside the anisotropic slab, with a direction given by the
resonance cone angle $\theta_c$, and then re-radiate into the
backward quadrant at the interface between the anisotropic waveguide
and the silicon slab to form a focal point at a certain distance from
the interface.
Figure~\ref{fig:RealWaveguide}\subref{fig:SiO2.Ag_LensImage} shows
the normalized amplitudes of magnetic and electric fields at the
image plane. The fundamental difference between the present imaging
system and the anisotropic bilayer lens \cite{ Liu_OL_2009} is that
the bilayer lens preserves subwavelength information through
continuous spectra, whereas the present system employs only a
discrete number of leaky modes to accomplish resolution enhancement.

\vspace{-2mm}

\bibliographystyle{ol}

\begin{thebibliography}{10}
\newcommand{\enquote}[1]{``#1''}

\bibitem{Oliner_LeakyWaves}
A.~A. Oliner and D.~R. Jackson, in \emph{Antenna Engineering
Handbook}, J.~L.
  Volakis, ed. (McGraw-Hill, New York, 2007), 4th ed.

\bibitem{Enoch_PRL_2002}
S.~Enoch, G.~Tayeb, P.~Sabouroux, N.~Gu\'{e}rin, and P.~Vincent,
Phys. Rev.
  Lett. \textbf{89}, 213902 (2002).

\bibitem{Baccarelli_IEEETMTT_2005}
P.~Baccarelli, P.~Burghignoli, F.~Frezza, A.~Galli, P.~Lampariello,
G.~Lovat,
  and S.~Paulotto, IEEE Trans. Microwave Theory Tech. \textbf{53}, 32 (2005).

\bibitem{Alu_IEEETAP_2007}
A.~Al\`{u}, F.~Bilotti, N.~Engheta, and L.~Vegni, IEEE Trans.
Antennas Propag.
  \textbf{55}, 882 (2007).

\bibitem{Itoh_Metamaterials}
C.~Caloz and T.~Itoh, \emph{Electromagnetic Metamaterials:
Transmission Line
  Theory and Microwave Applications} (Wiley, New York, 2005).

\bibitem{Smith_PRL_2003}
D.~R. Smith and D.~Schurig, Phys. Rev. Lett. \textbf{90}, 077405
(2003).

\bibitem{Liu_OL_2009}
H.~Liu, Shivanand, and K.~J. Webb, Opt. Lett. \textbf{34}, 2243
(2009).

\bibitem{Engheta_Science_2007}
N.~Engheta, Science \textbf{317}, 1698 (2007).

\bibitem{Podolskiy_PRB_2005}
V.~A. Podolskiy and E.~E. Narimanov, Phys. Rev. B \textbf{71},
201101 (2005).

\bibitem{Palik_Optical}
E.~D. Palik, ed., \emph{Handbook of Optical Constants of Solids}
(Academic
  Press, New York, 1998).

\end{thebibliography}

\end{document}